\documentclass[prl,twocolumn]{revtex4}
\usepackage{graphicx,amssymb,amsmath}
\usepackage{hyperref}
\usepackage{enumitem}
\usepackage{empheq}
\usepackage{color,soul}
\usepackage{mathrsfs}
\usepackage{afterpage}
\usepackage[caption=false]{subfig}
\usepackage{url}

\begin{document}

\title{Critical factors for mitigating car traffic in cities}

\author{Vincent Verbavatz}
\affiliation{Institut de Physique Th\'eorique, Universit\'e Paris Saclay, CEA, CNRS, F-91191 Gif-sur-Yvette, France}

\author{Marc Barthelemy}
\email{Corresponding author. Email: marc.barthelemy@ipht.fr}
\affiliation{Institut de Physique Th\'eorique, Universit\'e Paris Saclay, CEA, CNRS, F-91191 Gif-sur-Yvette, France}
\affiliation{Centre d'Analyse et de Math\'ematique Sociales, (CNRS/EHESS) 54, Boulevard Raspail, 75006 Paris, France}

\begin{abstract}

  Car traffic in urban systems has been studied intensely in past decades but models are either limited to a specific aspect of traffic or applied to a specific region. Despite the importance and urgency of the problem we have a poor theoretical understanding of the parameters controlling urban car use and congestion. Here, we combine economical and transport ingredients into a statistical physics approach and propose a generic model that predicts for different cities the share of car drivers, the $CO_2$ emitted by cars and the average commuting time. We confirm these analytical predictions on 25 major urban areas in the world, and our results suggest that urban density is not the most relevant variable controlling car-related quantities but rather are the city's area size and the density of public transport. Mitigating the traffic (and its effect such as $CO_2$ emissions) can then be obtained by reducing the urbanized area size or, more realistically, by improving either the public transport density or its access. In particular, increasing the population density is a good idea only if it also increases the fraction of individuals having access to public transport.

\end{abstract}

\maketitle

\section{Introduction}

As most humans now live in urban areas and two-third of the world population will live in cities by 2050 \cite{UN}, understanding urban mobility patterns \cite{Barbosa:2018} has become paramount in reducing transport-related greenhouse gas emissions and crucial to efficient environmental policies \cite{Dodman:2009,Glaeser:2010,Newman:2006,Oliveira:2014}. In a seminal paper, Newman and Kenworthy correlated transport-related quantities (such as gasoline consumption) with a determinant spatial criterion: urban density \cite{Newman:1989}. Higher population density areas were shown to have reduced gasoline consumption per capita and thus reduced gas emissions. Their result had a significant impact on urban theories over the last decades and has become a paradigm of spatial economics \cite{Creutzig:2015}. This study is however purely empirical and has no theoretical foundation, which casts some doubts about the importance of density as the sole determinant of gasoline consumption and other car dependent quantities.

On the other hand, car traffic has been studied at various granularities -- and mostly at the theoretical level of a single lane -- with various tools ranging from agent-based modelling (see for example \cite{Balmer:2004,Bazzan:2014}), cellular automata \cite{Nagel:1992}, or hydrodynamic approaches (see for example the review \cite{Nagatani:2002} on various physics type approaches).  Car congestion was also considered from the point of view of economics with discussion about its determinants (see for example 
\cite{Arnott:1990,Duranton:2011,Couture:2018,Duranton:2019}). On a more applied side, traffic forecast models such as the four-step model and variants (see for example the handbook \cite{Hensher:2007} and references therein) are used for key purposes in transportation policy and planning. These models rely on data about a specific area such as population, employment, actual traffic, etc., and are used to forecast the traffic on a given infrastructure (or a project) and its environment impact. In all cases, these models are fine-tuned for specific areas and we don't have so far a generic model that is able to predict the value of traffic-related quantities for any city, and to point to the critical parameters and dominant mechanisms of traffic in urban areas. In this paper, we address this problem and propose a theoretical model of urban daily commuting relying on two crucial ingredients: the coexistence of cars and mass rapid transit (MRT), and traffic congestion. Combining these ingredients within a disordered system type approach allows us to derive significant conclusions concerning MRT transport ridership and transport-related greenhouse gas emissions. We compare our predictions to empirical data obtained for 25 major cities in the world (see Sup. Mat. for details about data) showing an excellent agreement given the simplicity of the model and the absence of any ajustable parameter. In this approach we deliberately left out details of these systems and focused on the basic processes that capture the complexity of urban systems while accounting for qualitative and quantitative behaviors \cite{Pumain:2004,Gonzalez:2008,Bettencourt:2010,Batty:2008,Barthelemy:2016}.

\section{The model}

\subsection{Symplifying the Fujita-Ogawa model}

According to the classical urban ecomomics model of Fujita and Ogawa \cite{Fujita:1982}, individuals choose job and dwelling places that maximize their net income after deduction of rent and commuting costs. More precisely, an agent will choose to live in $x$ and work at location $y$ such that the quantity
\begin{align}
Z(x,y)=W(y)-C_R(x)-G(x,y)
\end{align}
is maximum. The quantity $W(y)$ is the typical wage earned at location $y$,  $C_R(x)$ is the rent cost at $x$, and $G(x,y)$ is the generalized transportation cost to go from $x$ to $y$. There is also a similar equation for the profit of companies (that they want to maximize) and that we do not need here: we assume that employment is located at a unique center $y=0$ and that wages and rent costs are of the same order for all individuals. Most large cities are polycentric \cite{Louf:2013,Louf:2014} with the existence of many different activity centers but this first approach assumes that polycentricity does not change the order of magnitude of commuting trip distances (see also \cite{Louf:2014} on this point). We also assume that the residence location $x$ is given and random -- residence choice is obviously a complex problem and replacing a complex quantity by a random one is a typical assumption made in the statistical physics of complex systems. Within these assumptions, we obtain a simplified Fujita-Ogawa model where we focus on the mode choice: individuals have already a home and work at the CBD, and the problem is about choosing a transport mode. Within these assumptions, the maximization of $Z(x,0)$ implies the minimization of the transport cost $G(x,0)$: $\max Z(x,0) \Rightarrow \min G(x,0)$. Thus, individuals simply tend to choose the mode that minimizes their commuting costs to go the office located at $0$. In this perspective, we assume that individuals change more frequently their jobs than their homes. More generally, the Fujita Ogawa model gives a density that results of the interplay between residential and commuting costs and due to our simplifications, the density is now an exogenous quantity. 

\begin{figure}[!h]
\centering
\includegraphics[scale=0.4]{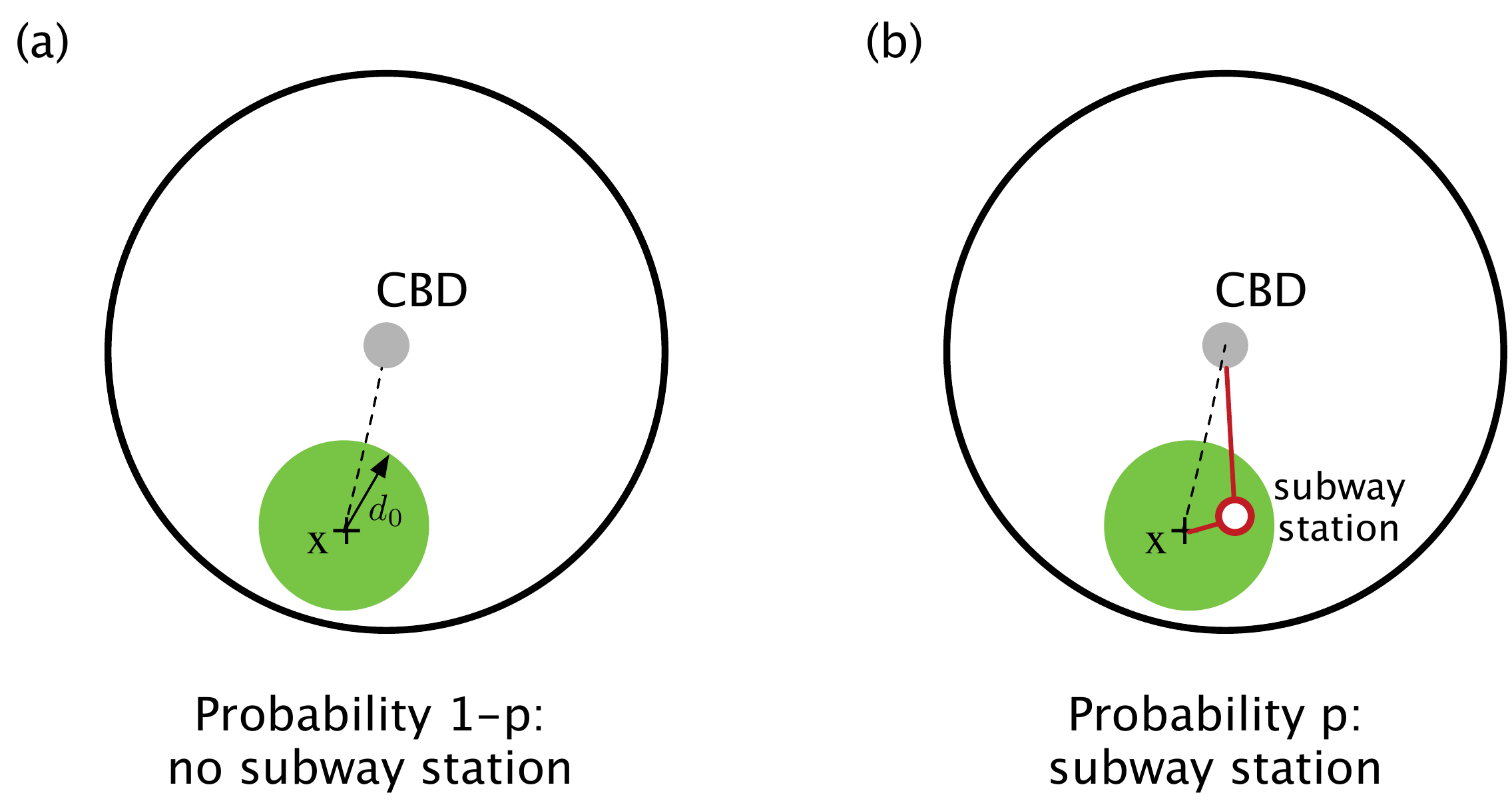}
+\caption{Sketch of the model. (a) For a given agent located at a random location $x$, there is no subway station located at a distance less than $d_0$ with probability $1-p$ (in the data used here $d_0=1km$). In this case the journey to work located at the central business district (CBD) is made by car (dashed line). (b) With probability $p$ there is a subway station in the neighborhood of $x$ and the agent has to compare the cost $G_{car}$ of car (dashed line) and the cost  $G_{MRT}$ of MRT (the trip is depicted by the red line) in order to choose the less costly transportation mode to go to the CBD.}
\label{fig:model}
\end{figure}

In order to discuss commuting costs, we assume that a proportion $p$ of the population has access (meaning having to walk less than 1km) to mass rapid transit such as the subway or elevated rail (we neglect buses or tramways here) whereas a share $1-p$ of the population has no choice but to commute by car (we assume that all individuals can drive a car if needed). In other words, $p$ is the probability to have access to the MRT (see Fig.~\ref{fig:model}), but it doesn't imply that it is the mode chosen: if the agent has access to the subway, he needs to compare the costs $G_{car}$ and $G_{MRT}$ in order to choose the lest costly transportation mode.

The existence of a single central business district, the location of homes, and the density of MRT are considered as exogenous variables. The important endogenous variable is here the share of car users and the time spent in traffic jams (allowing then to estimate CO2 emissions). All the assumptions used in this model are of course approximations to the reality but we claim here that our model captures the essence of the traffic in large urban areas phenomenon. Starting with a model containing all these various parameters would actually be not tractable and would hide the critical ingredients.

\subsection{Generalized cost}

In order to define the model completely we have to specify the expressions for the generalized costs. 
We will omit all other forms of commuting (walking, cycling, etc.), and we neglect spatial correlations between the densities of public transport and residence or population, which is an important assumption that certainly needs to be refined in future studies. The fraction $p$ of individuals that have a choice between car and MRT will choose the transport mode with the lowest generalized cost which takes into account both monetary costs and trip duration (see for example \cite{Crozet:2005,Glaeser:2008}). 

For cars, we include congestion described by the Bureau of Public Roads function (see \cite{Branston:1976} and Methods) which captures the main effect: increasing the traffic on a road will decrease the effective speed on it. For the MRT, we neglect its monetary cost which is small in comparison with the one for cars. We then obtain the corresponding generalized costs for cars and the MRT under the form
\begin{align}
&G_{car}(x)=C_c+\frac{d(r)}{v_c}V\left(1+\left(\frac{T}{c}\right)^{\mu}\right)\\
&G_{MRT}(x)=V\left(f+\frac{d(r)}{v_m}\right)
\end{align}
where $C_c$ is the daily cost of a car, $v_c$ and $v_m$ are respectively the car and MRT velocities, $c$ is the road capacity of the city, $f$ the walking plus waiting time for transit, $d(r)$ the distance between home located at a distance $r$ from the central business district, $T$ the total car traffic, and $\mu$ the exponent that characterizes the sensitivity to traffic. The quantity $V$ is the value of time defined in transport economics as the money amount that a traveler is willing to pay in order to save one hour of time. It is an increasing function of income and is bounded by the hourly wage. We assume here that driving is faster than riding public transport ($v_c>v_m$) but is more expensive. Once an individual has chosen a mode he sticks to it and will not reconsider his choice even if the traffic evolves. In other words, we assume here that individual habits have a longer time scale than traffic dynamics.

\section{Results}

\subsection{Critical distance}

\begin{figure}[!h]
\centering
\includegraphics[scale=0.4]{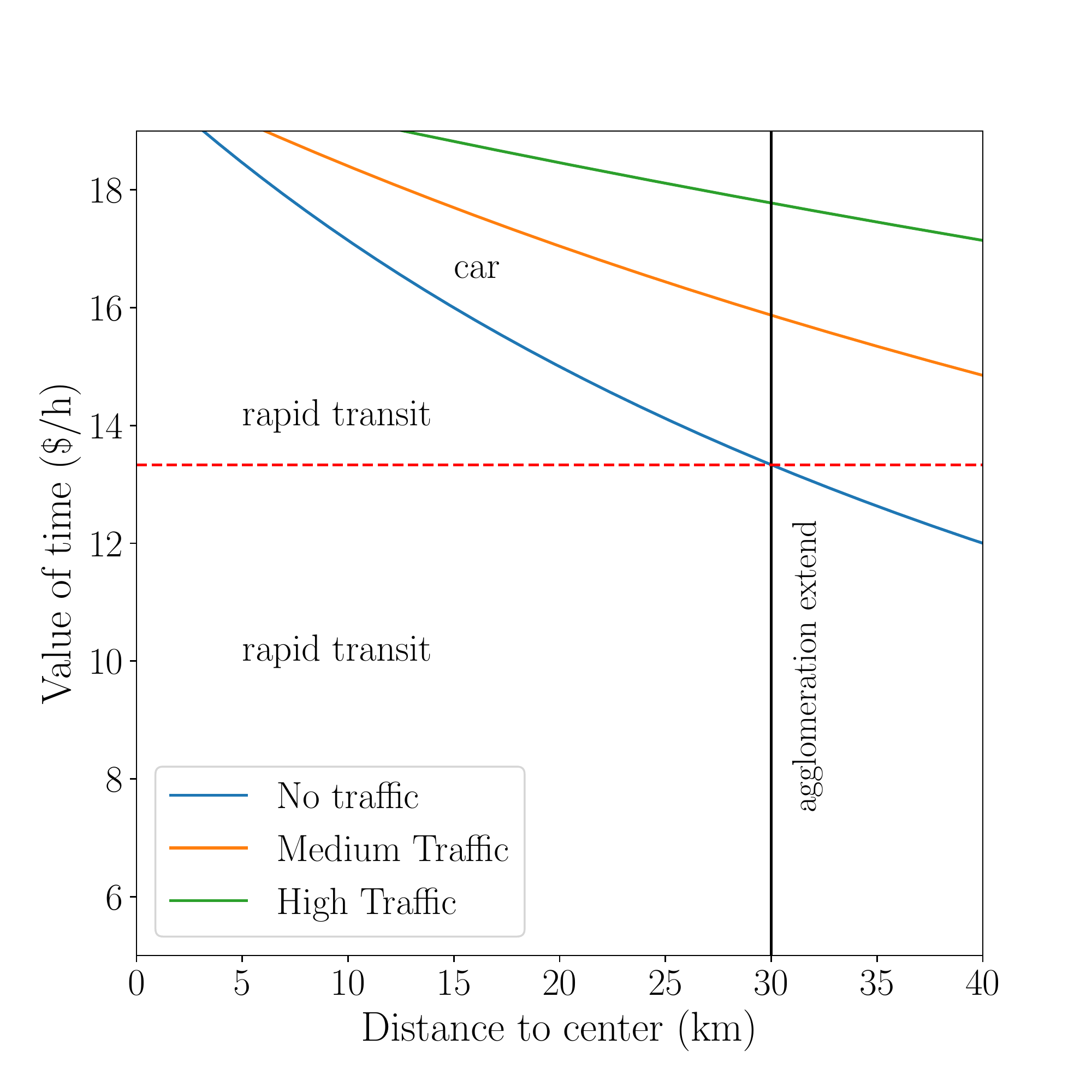}
+\caption{The most advantageous mode of transport depends on the value of time of individuals and the distance to the urban center. The limit between the two areas evolves with congestion: the larger the traffic (curves from blue to green) and the larger the area in which rapid transit is beneficial compared to car driving. The grey solid vertical line corresponds to the size of the urban area and indicates the critical value of time  (dashed red line) below which rapid transit is advantageous in the whole agglomeration whatever the value of congestion. The values of the parameters are chosen here as:  $C_c=15$~\$, $v_c=40$~km/h, $v_m=30$~km/h, $f=30$~minutes (see Material and Methods for a precise description of all data).}
\label{fig:1}
\end{figure}

Individual mobility is then governed by comparing these costs $G_{car}$ and $G_{MRT}$ and will depend on exogenous parameters such as car and subway velocities, car costs, etc.). In the general mode choice theory (see for example \cite{BenAkiva:1985}), given the values of
the costs $G_{car}$ and $G_{MRT}$ there is a probability $P_C=F(G_{car}-G_{MRT})$ to choose the car. The function $F$ is in general
smooth and satisfies $F(-\infty)=1$ and $F(+\infty)=0$ and we consider here the simplest case where $F(x>0)=0$ and $F(x<0)=1$. An individual
located at $x$ with access to the MRT will then choose to use the car if $G_{car}<G_{MRT}$ which implies a condition on the value of time of
the form $V<V_m(r)$ where $V_m(r)$ depends on the parameters of the system
and on $d(r)$ and reads
\begin{align}
V_m(r)=\frac{C_c}{f+d(r)\left(\frac{1}{v_m}-\frac{1}{v_c}\left(1+\left(\frac{T}{c}\right)^\mu\right)\right)}
\end{align}
Also, writing the equality $G_{car}=G_{MRT}$ between generalized costs of car and MRT leads to the critical distance given by
\begin{align}
d(V,T)=\min\left(L,
\frac
{\frac{C_c}{V}-f}
{
\frac{1}{v_m}-\frac{1}{v_c}(1+(\frac{T}{c})^\mu)
}
\right)
\label{eq:dvt}
\end{align}
where $L \sim \sqrt{A}$ ($A$ is the area size of the city) is the largest extent of the city. This critical distance evolves as traffic increases (Fig.~\ref{fig:1}), translating the fact that driving is less advantageous when the traffic is large. This expression also shows that individuals with a small value of time are more likely to use public transport, since they are more apt to spend time than money. Distance to the center is pivotal in this decision process: too far from the center - further than a critical distance $d(V, T)$ - individuals favor driving to avoid lengthy journeys, and the richer they are, the smallest this distance.

Writing $d(V, T=T^*)= L$, gives
the critical maximal traffic $T^*$ for which driving is less beneficial in the whole
agglomeration:
\begin{align}
T^*=c\left[v_c\left(\frac{1}{v_m}-\frac{1}{v_c}-\frac{\frac{C_c}{V}-f}{L}\right)\right]^{1/\mu}
\end{align}

\subsection{Predicting the car share}

When population increases, car traffic has essentially two sources: first, individuals may not have access to the
MRT and second, if they have access to it, they might be too far and will prefer to take the car. This can be summarized by the following differential equation for the variation of car traffic T when P varies 
\begin{equation}
  \frac{\mathrm{d}T}{\mathrm{d}P}=1-p+p\left[1-\eta(d(V,T))\right]
  \label{eq:diff}
\end{equation}
where the first term of the r.h.s. corresponds to the $1-p$ share of individuals far from MRT stations, and the second term to the fraction $1-\eta$ of individuals that have access to the MRT and are living further from the center than $d(V,T)$. This equation is valid for $T<T^*$ and for $T>T^*$ we have $\eta=1$. By considering statistically isotropic cities with population density $\rho(r)$ ($r$ is the distance to the CDB), we obtain
\begin{align}
\eta(d(V,T))=\frac
{\int_0^{d(V,T)}\mathrm{d}rr\rho(r)}
{\int_0^{L}\mathrm{d}rr\rho(r)}
\end{align}

We have to plug in this expression together with Eq.~\ref{eq:dvt} of $d(V,T)$ and to solve the differential equation Eq.~\ref{eq:diff} which in general is too difficult. The density is endogenous in the original Fujita-Ogawa model but due to our simplifications is now an exogenous quantity. In general the density is decreasing with $r$ and a simple choice is $\rho(r)=\rho_0\mathrm{exp}(-r/r_0)$ and we then obtain at dominant order in $T/C$
\begin{align}
  \frac{\mathrm{d}T}{\mathrm{d}P}\approx 1-p\mathrm{e}^{-r/r_0}\left(1+\frac{a}{r_0}\right)+{\cal O}\left(\left(\frac{T}{C}\right)^\mu\right)
  \label{eq:infra}
\end{align}
where $a=(C_c/V-f)/(1/v_m-1/v_c)$. We note here that in the uniform density case $\rho(r)=\rho_0$, we obtain $\eta=\pi d(V,T)^2/A$ and  $T\simeq\left(1-\frac{p}{Ab^2}\right)P+{\cal O}\left(\left(\frac{T}{c}\right)^\mu\right)$ where $b$ is a function of the exogenous parameters $b=(1/v_m-1/v_c)/\sqrt{\pi}(C_c/V-f)$. 

This result Eq.~\ref{eq:infra} is valid 
For $P < P^*=T^*/\left(1-p\mathrm{e}^{-a/r_0}(1+a/r_0)\right)$ since for $d(V,T)>L$ the only source of car traffic comes from individuals who do not have access to the mass rapid transit, which leads to  $dT/dP=1-p$ implying
\begin{equation}
T=(1-p)(P-P^*)+T^*
\end{equation}
We note that even if this result seems somewhat simple, it derives from non-trivial considerations such as the comparison of the critical distance and the area size. The important fact to retain here is that cost considerations in the case where a mode choice is available usually lead for large urban areas to leave the car and take the MRT. Also, we note that the effect of the density is essentially on the first regime where $T<T^*$ and therefore does not impact our conclusions. 

\begin{figure}[!h]
\centering
\includegraphics[scale=0.3]{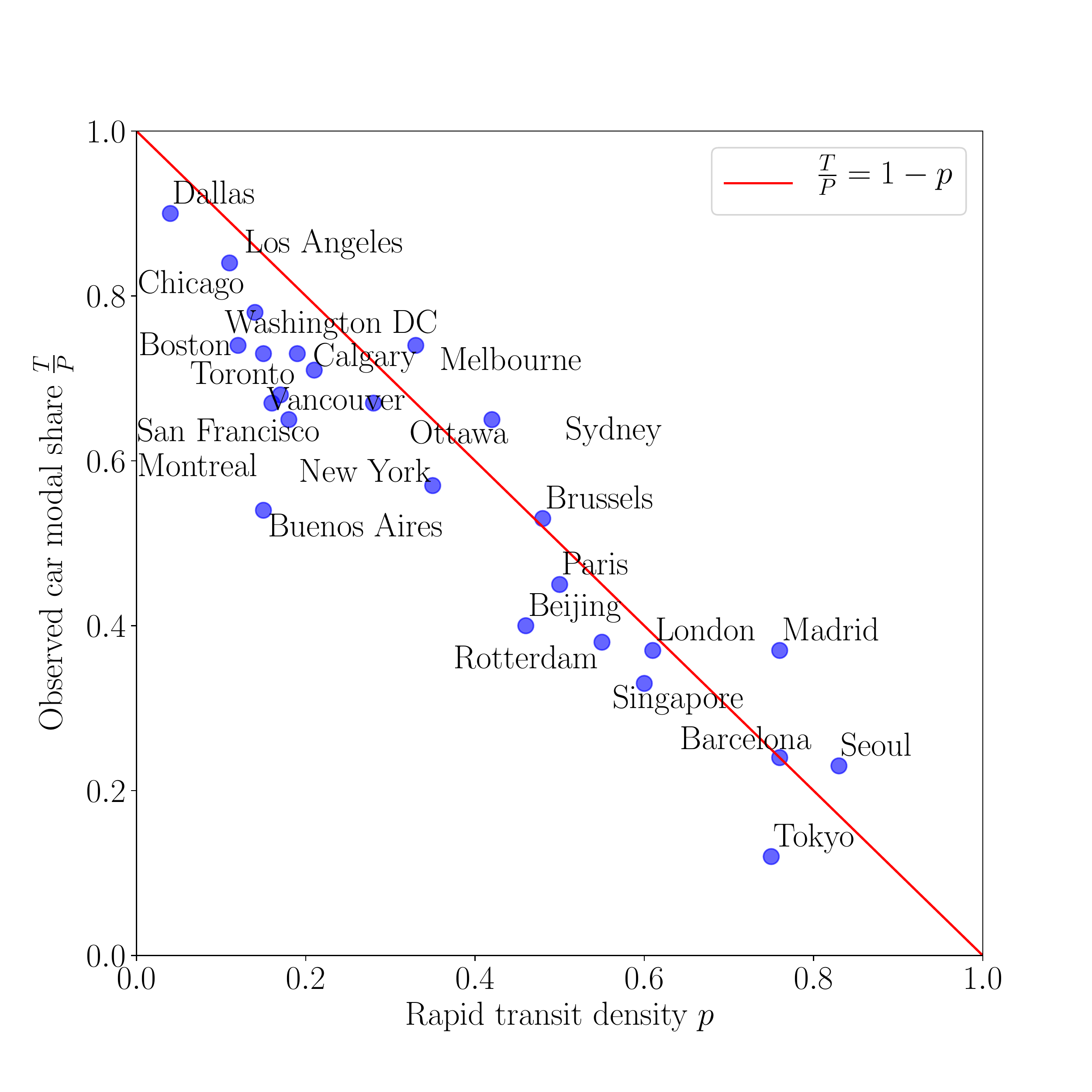}
\caption{Comparison between the observed car modal share $T/P$ and the share of population $p$ living near rapid transit stations (less than 1 km) for 25 metropolitan areas in the world. The red line is the prediction of our model ($R^2=0.69$). Given the absence of any tunable parameter the agreement is satisfactorily, and discrepancies are probably mostly due to the existence of other modes of transport (walking or cycling), lower car ownership rates, or a higher cost of the MRT, etc.}
\label{fig:1-p}
\end{figure}

Compiling data from 25 megacities in the world (see Material and Methods) for which we found an estimate of the population having access to the MRT (and therefore the quantity $p$), we compute the critical levels $T^*$ and $P^*$ (we note that for all these cities the definition of $p$ is the same: it is the
share of individuals living within $1$km from a MRT station). For reasonable values of time \cite{Numbeo:2018,Crozet:2005}, most of cities have $T^*=0$ and all have $T^*\ll P$ ($T^*$ ranges from 0 to 50\% of $P$). Cases where $T^*=0$ mean that even at zero traffic the critical distance $d(V,T=0)$ is larger than the typical size of the city $L$. This depends a bit on the value of time but in most
cases we do observe small values of $T^*$ indicating that cities are
mostly in the saturated regime where the MRT is always more advantagous than the car. Public transports are so economical (compared to cars) that people living near rapid transit stations are highly likely to ride them. Thus, traffic does not appear as a determinant parameter in individual mobility choices as it concerns mostly individuals who have no choice but to drive and who suffer from onerous commuting costs and unavoidable time-consuming trips as traffic increases. We note here that correlations between the neighborhoods with low MRT density (small $p$) where individuals have expensive commuting costs and their revenue is an interesting field of study and appears as an important source of urban segregation \cite{Glaeser:2008}. Since we have in general $T^*, P^*\ll P$, we obtain the simple prediction that $\frac{T}{P}\simeq 1-p$, a non-trivial consequence of rapid transit cheapness and individual choices of mobility: we recall that $p$ is the probability to leave in the vicinity of a MRT station but that doesn't imply the use of this transport mode. We compare the empirical car modal shares $\frac{T}{P}$ of these cities (see data description in Material and Methods) to our prediction on Figure \ref{fig:1-p} and observe -- considering the simplicity of the model  -- a very good agreement and a relevant linear trend highlighting the efficiency of public transportation in reducing traffic. 
In particular, most of the European cities are well described by our prediction and we observe a few deviations. These discrepancies can probably find their origin in the existence of other modes of commuting, lower car ownership rates (e.g. in Buenos Aires \cite{Roque:2016}), lower road capacities, higher cost of MRT, or a high degree of polycentrism.

\subsection{Estimating the emitted CO$_2$}

\begin{figure}
\centering
\includegraphics[scale=0.3]{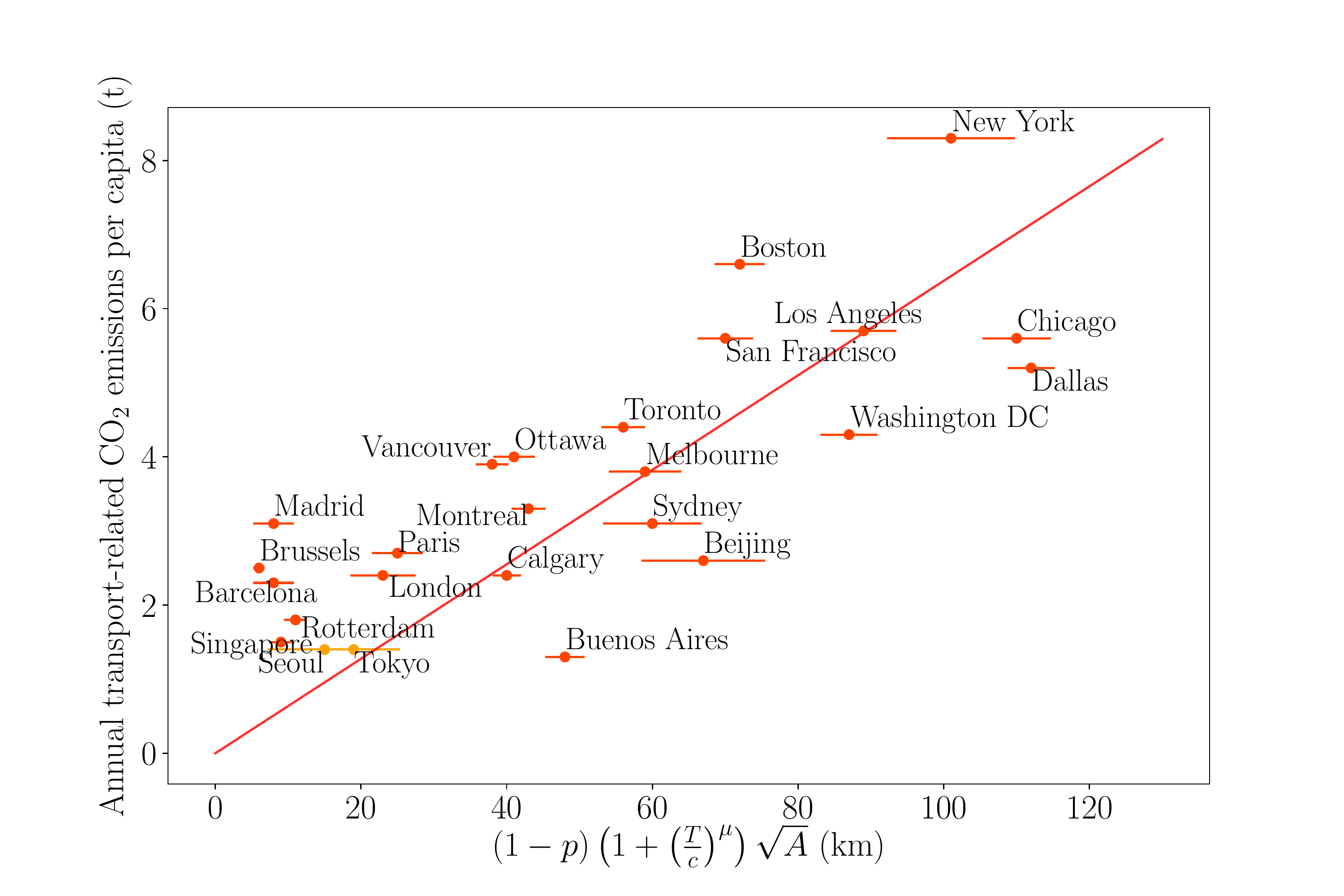}
\caption{Comparison between the annual transport-related CO$_2$ emissions per capita and the effects of congestion, area size and rapid transit density predicted by our model.
The red line is the linear fit of the predicted form $y=\alpha x$ where $\alpha\approx 0.064$~CO$_2$tons/km/hab/year (the Pearson coefficient is $0.79$)
We had no congestion estimate for Seoul and Tokyo and we used an average congestion rate $\tau=50\%$. The error bars are computed for a typical error of $10\%$ on $p$ and $\tau$.}
\label{fig:CO2}
\end{figure}

Our model also provides a prediction for the transport-related gas emissions and we will focus on the $CO_2$ case for which we obtained data. We make the simplest assumption where these emissions are proportional to the total time spent on roads. The quantity of $CO_2$ emitted for a driver residing at $x$ is then given by $Q_{CO_2}(r)\propto d(r)[1+(T/c)^\mu]$ which leads to a total 
\begin{align}
Q_{CO_2}&\propto\sum_{drivers\; i}d(r_i)\left[1+\left(\frac{T}{c}\right)^\mu\right]\\
&\propto g\sqrt{A} (1-p)P\left[1+\left(\frac{T}{c}\right)^\mu\right]
\end{align}
where we used our result for the total traffic $T=(1-p)P$. We assumed that the sum $\sum_i d(r_i)$ is of the form $g\sqrt{A}T$ where $\sqrt{A}$ sets the scale of displacement and where the prefactor $g$ encodes the geometrical aspect of car mobility in the city, including the spatial distribution of residences and activities, and the transport infrastructure. Its estimation probably requires a more detailed, specific calculation but the important aspect here is the scaling with $\sqrt{A}$ (see also the Supp. Info.). We thus obtain that the annual CO$_2$ emitted by car and per capita is given by
\begin{align}
\frac{Q_{CO_2}}{P}\propto \sqrt{A}(1-p)(1+\tau)
\label{eq:qco2}
\end{align}
 where $\tau=\left(\frac{T}{c}\right)^{\mu}$ is the average delay due to congestion and is empirically accessible from the TomTom database \cite{tomtom:2008} (while the capacity $c$ is usually not empirically accessible). It is interesting to note that $Q_{CO_2}$ is then the product of three main terms: size of the city $\times$ fraction of car drivers $\times$ congestion effects, which correspond indeed to the intuitive expectation about the main ingredients governing car traffic. Also, we note here that the dominant term of $Q_{CO_2}$ is proportional to the population indicating a simple linear scaling, and that nonlinear term could possibly appear as corrections, in contrast with previous results \cite{Bettencourt:2007,Louf:2014}, and which could explain the difficulties to reach a consensus about the behavior of this type of quantities (see for example \cite{Leitao:2016}). We compare our prediction Eq.~\ref{eq:qco2} to disaggregated values of urban CO$_2$ emissions on Figure \ref{fig:CO2} and observe a good agreement. We observe some outliers like Buenos Aires which has a very small car ownership rate and thus lower than expected CO$_2$ emissions and areas such as New York which appears to be one of the largest transport CO$_2$ emitter in the world \cite{oecd:2016}.
This result illustrates the role played by public transport and traffic in modulating transport-related CO$_2$ emissions. Most importantly, we identify urban sprawl ($\sqrt{A}$) as a major criterion for transport emissions. We note that if we introduce the average population density $\rho=\frac{P}{A}$, we can rewrite our result as
\begin{align}
\frac{Q_{CO_2}}{P}\propto \frac{\sqrt{P}}{\sqrt{\rho}} (1-p)(1+\tau)
\end{align} 
i.e. $\frac{Q_{CO_2}}{P} \propto \rho^{\frac{-1}{2}}$ since $\sqrt{P}$ is a slowly-varying function within the scope of large urban areas. We understand here how Newman and Kenworthy \cite{Newman:1989} could have obtained their result by assuming the density to be the control parameter.  However, even if fitting data with a function of $\rho$ is possible, our analysis shows that it is qualitatively wrong: the area size $A$ and the public transport density $p$ seem to be the true parameters controlling car-related quantities such as $CO_2$ emissions. Mitigating the traffic is therefore not obtained by increasing the density but by reducing the area size and improving the public transport density. Increasing the population at fixed area would increase the emission of $CO_2$ (due to an increase of traffic congestion leading to an increase of $\tau$) in contrast with the naive Newman-Kenworthy assumption where increasing the density leads to a decrease of $CO_2$ emissions.

We also note that
\begin{align}
\frac{\partial\log Q_{CO_2}}{\partial \log p} = -\frac{p}{1-p}
\end{align}
can be relatively large (in absolute value) while
\begin{align}
\frac{\partial\log Q_{CO_2}}{\partial\log A} = \frac{1}{2}
\end{align}
is small which suggests that the increase of transport density is in general more efficient than an area decrease (which is also less feasible). We also understand that if mass rapid transit is efficient in reducing transport-related greenhouse emissions, on the other hand reducing the road capacity $c$ (to deter individuals from driving) may not be a good remedy, as we identified drivers with individuals with no other solution. Worse, by making their trips longer, diminishing c indirectly contributes to higher emissions and pollution as well as potentially socially segregating situations.

\subsection{Average commuting time}

Finally, we can also estime the average commuting time (details are given in Material and Methods) and we obtain for the one-trip commuting time $\tau_c$ averaged over the population the following expression
\begin{align}
\overline{\tau_c}=p\left(f+\frac{g\sqrt{A}}{v_m}\right)+(1-p)\frac{g\sqrt{A}}{v_c}(1+\tau)
\label{eq:tau}
\end{align}
where $g$ is a geographical factor that encodes the spatial complexity of trips. The comparison with the data displays relatively large fluctuations, but our analysis seems to capture the main trend and the one-parameter fit over $g$ using our expression Eq.~\ref{eq:tau} leads to the average value $g\approx 0.203$. We note that for a uniform distribution of residences, a simple calculation leads to $g=2/3\sqrt{\pi}\approx 0.376$ and in the simple isotropic case where the density decreases with the distance $r$ to the center as $\rho(r)=\rho_0(1-r/L)$, we obtain $g=1/3\sqrt{\pi}\approx 0.188$. The average value $g$ obtained by the fit is in between these two theoretical estimates. However, polycentrism and more generally the spatial organization of the city and its infrastructure certainly play an important role here and our model can only provide a first approximation to the average commuting time.

\section{Discussion}

We presented a parsimonious and generic model for the car traffic and its consequences in large cities, and tested against empirical data for 25 large cities in the world. This approach illustrates how a combination of statistical physics, economical ingredients and empirical validation can lead to a robust understanding of systems as complex as cities. In particular, this approach is in contrast with the commonly accepted idea that urban density is pivotal, and our aim here was to capture the essence of the urban mobility phenomenon and to obtain analytical results for the car traffic and the quantity of emitted $CO_2$. Our analysis shows that traffic related quantities are governed by three factors: access to mass rapid transit, congestion effects and the urban area size. In order to reduce $CO_2$ emissions for example, our model suggests to increase public transport access either by increasing the density around MRT stations or to increase the density of public transport (in contrast with the conclusions of an econometric study in the US \cite{Duranton:2011}), or to reduce the urban area size (impossible in most contexts). Increasing the cost of car use seems actually unable to lower car traffic in the absence of alternative transportation means.  Also, it is important to note that increasing in general the density at fixed both area size and MRT access would actually increase $CO_2$ emissions.  Finally, we insist on the fact that in this model we voluntarily left out a number of parameters such as other transportation modes (buses, tramway), polycentrism, the transport network structure, fuel price and tax, dynamic road pricing, etc. but our main point was to fill a gap for understanding traffic in urban areas by proposing a parsimonious model with the smallest number of parameters and the largest number of predictions in agreement with data. Given the simplicity of this model we cannot expect a perfect agreement with data for various and different cities, but it seems that this approach captures correctly all the trends and identifies correctly the critical factors for car traffic. This seems to be a basic requirement before adding other factors and increasing the complexity of the model. Also, it seems at this stage necessary to first encourage the measure and sharing of data such as the density of public transport in order to propose further tests of our theoretical framework.

\section*{Material and Methods}

\subsection*{Data sources}

We studied 25 metropolitan areas from Europe, America, Asia and Australia. The number of cities was limited by the availability of data on MRT accessibility and reliable modal share estimates. All the data is 
freely available and we list here their sources. We also provide a file together with this submission with all the data used in this study (see Supp. Info. file).

\begin{itemize}

\item{} The definition of a metropolitan area varies from one country 
to another but we aimed at assembling a statistically coherent set of 
agglomerations. Populations and areas were collated from Wikipedia data on metropolitan statistical areas in accordance with national definitions.

\item{} Various indicators were compiled from diverse sources: the
  quantity $p$ is taken from transit accessibility reports
  \cite{Marks:2016,Singer:2014,IBSA:2015,Singa:2013,Japan:2013,chart}. A
  common metric is the share of individuals living within 1~km from a MRT station (this
  is the assessed maximal walking distance). We note however that this
  is an unusual indicator of mobility that is  not always easy to find
  and was here the main bottleneck for getting a larger set of
  cities. Modal shares were obtained from
  \cite{McKenzie:2015,Canada:2017,Kennis:2016,chart,EMTA:2012,ATM,singa2},
  CO$_2$ emissions from \cite{oecd:2016,WB1,WB2,GHC,WRI}, and
  commuting time from
  \cite{Lebrun,cana:2016,uranga,ipea,acs,bitre,LIT,Korea,beijing}.

\item{} Values of time were assessed by taking half of the hourly wage after tax for each city \cite{Numbeo:2018, Crozet:2005}.  

\item{} Congestion delays were taken from the TomTom index \cite{tomtom:2008}, except for Seoul and Tokyo where we had no data (we assumed an average delay of $50\%$).  

\item{} The road capacity was computed from the congestion delay
  $\tau$ by $c\sim P/\tau^{1/\mu}$ with the value $\mu=2$ 
that we used for all cities.

\item{} For the velocities $v_m$ and $v_c$ and the costs $C_c$ and $f$
  we choose the same values for all cities. For $v_m$ wikipedia data
  \cite{metro_speeds} displays values in the range $25-35~$km/h depending on the city and we
took $v_m\approx$~30km/h. The free flow car velocity $v_c$ also depends a bit on the
city and varies from $30$~km/h (without congestion effect) in European cities such as Paris to $56$~km/h
in some american cities \cite{vcsite}. We decided to take an average value
$v_c\approx 40$~km/h.

\item{} For $C_c$, we used a cost simulator \cite{calc} which gives on average a value of $C_c\approx 15$ USD per trip. 

\item{} For $f$ we counted on average $10$ minutes to reach the station, $10$ minutes to reach
the office and $10$ minutes for transit for a total of about $f\approx
30$ minutes.  This value can certainly be improved but it is difficult to get and we
expect it not to vary too much (at least not over more than one order
of magnitude).

\end{itemize}

\subsection*{Average commuting time}

\begin{figure}
\centering
\includegraphics[scale=0.3]{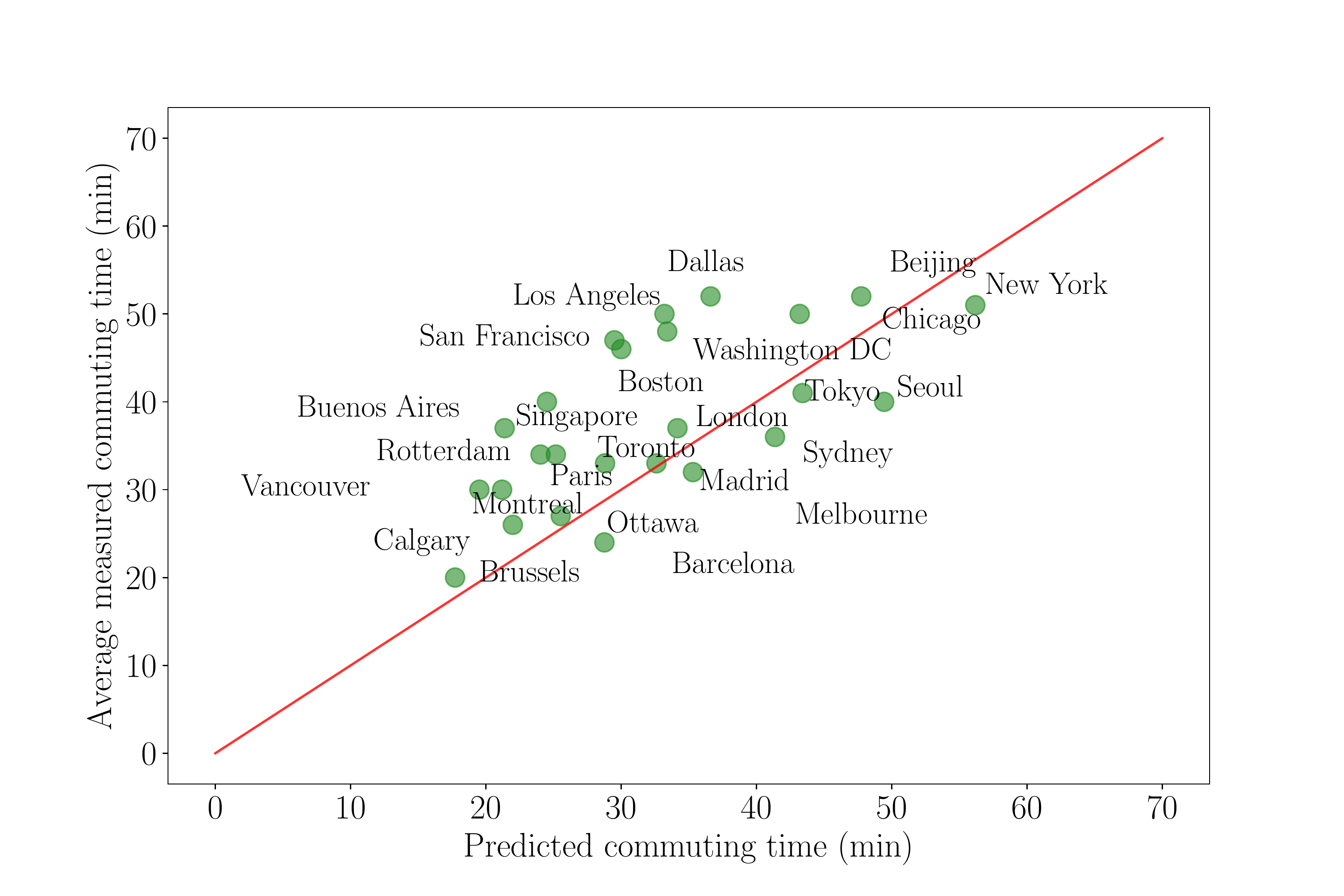}
\caption{Average commuting time $\overline{\tau_c}$ measured for different cities versus the predicted value of our model (Eq.~\ref{eq:tau2}). We perform the fit on the single parameter $g$ using Eq.~\ref{eq:tau2}, leading to an average value $g\approx 0.203$ (the Pearson correlation is here $0.65$). The existence of large fluctuations around our prediction is probably connected to the spatial organization of the city and the structure of the transportation networks.}
\label{fig:time}
\end{figure}

The commuting time for MRT users is given by $f+g\sqrt{A}/v_c$ while for car users it is $g\sqrt{A}(1+\tau)/v_c$ where $g$ is a geographical factor. We therefore obtain for the one-trip commuting time $\tau_c$ averaged over the population the following expression
\begin{align}
\overline{\tau_c}=p\left(f+\frac{g\sqrt{A}}{v_m}\right)+(1-p)\frac{g\sqrt{A}}{v_c}(1+\tau)
\label{eq:tau2}
\end{align}
The comparison of this result with empirical data is shown on Figure \ref{fig:time}.
Even if we observe relatively large fluctuations, our analysis seems
to capture the main trend and the one-parameter fit using our
expression Eq.~\ref{eq:tau2} leads to the average value $g\approx
0.203$. We note that for a uniform distribution of residences, a
simple calculation leads to $g=2/3\sqrt{\pi}\approx 0.376$ and in the
simple isotropic case where the density decreases with the distance
$r$ to the center as $\rho(r)=\rho_0(1-r/L)$, we obtain
$g=1/3\sqrt{\pi}\approx 0.188$. The average value $g$ obtained by the
fit is in between these two theoretical estimates. From
Fig.~\ref{fig:time}  we can compute the `effective $g_{\mathrm{eff}}$'
for each city and compare it to the average value $g$. This quantity
$g_{\mathrm{eff}}$ (and the ratio $\eta=g_{\mathrm{eff}}/g$) encodes
both the complexity of the population and activity densities, and of
the transportation infrastructure. For our set of cities, the average
ratio is of order $1.37$ and we observe outliers with small values
such as Barcelona ($\eta=0.04$), Seoul ($\eta=0.62$), and large ones
(Buenos Aires $1.92$, Rotterdam $2.32$, Singapore $3.38$). Most of the
cities ($76\%$) have however here a ratio $\eta$ larger than $1$ which
implies that our model underestimates in general the commuting
distance. Many local effects can explain the variations observed in
the commuting distance: the existence of polycentricity could reduce
the commuting distance (for example, it seems that we overestimate the
commuting time for New York which might be a consequence of
polycentrism), but other factors such as poor transportation
infrastructures (or traffic bottlenecks due to geographical
constraints) could have the opposite effect of increasing this
quantity. In general we expect that the transport network structure
will have an important impact, especially if it is very anisotropic in
space. The determination of the commuting distance for each city
probably implies to take into account a large number of specific
details, but our analysis shows that it is consistent to assume that
the order of magnitude of this quantity is $\sqrt{A}$.

\section*{Acknowledgments}
 VV thanks the IPhT for support during his internship.

\end{document}